\documentclass[12pt]{article}
\usepackage{a4wide}
\usepackage{epsfig}
\usepackage{amsmath}

\newlength{\absize}
\catcode`@=11
\def\citer{\@ifnextchar
[{\@tempswatrue\@citexr}{\@tempswafalse\@citexr[]}}

\def\@citexr[#1]#2{\if@filesw\immediate
  \write\@auxout{\string\citation{#2}}\fi
  \def\@citea{}\@cite{\@for\@citeb:=#2\do
    {\@citea\def\@citea{--\penalty\@m}\@ifundefined
       {b@\@citeb}{{\bf ?}\@warning
       {Citation `\@citeb' on page \thepage \space undefined}}%
\hbox{\csname b@\@citeb\endcsname}}}{#1}}
\catcode`@=12

\begin{document}
  \thispagestyle{empty}
  \pagestyle{empty}
  \renewcommand{\thefootnote}{\fnsymbol{footnote}}
\newpage\normalsize
    \pagestyle{plain}
    \setlength{\baselineskip}{4ex}\par
    \setcounter{footnote}{0}
    \renewcommand{\thefootnote}{\arabic{footnote}}
\newcommand{\preprint}[1]{%
  \begin{flushright}
    \setlength{\baselineskip}{3ex} #1
  \end{flushright}}
\renewcommand{\title}[1]{%
  \begin{center}
    \LARGE #1
  \end{center}\par}
\renewcommand{\author}[1]{%
  \vspace{2ex}
  {\Large
   \begin{center}
     \setlength{\baselineskip}{3ex} #1 \par
   \end{center}}}
\renewcommand{\thanks}[1]{\footnote{#1}}
\begin{flushright}
\end{flushright}
\vskip 0.5cm

\begin{center}
{\large \bf Testing Abelian dyon - fermion  Bound System}
\end{center}
\vspace{1cm}
\begin{center}
Jian-zu Zhang$^{a,b, \S}$
\end{center}
\vspace{1cm}
\begin{center}
$^a$ Institute for Theoretical Physics, Box 316,
East China University of Science and Technology,
Shanghai 200237, P. R. China \\
$^b$  Department of Physics,
University of Kaiserslautern, PO Box 3049, D-67653  Kaiserslautern,
Germany 
\end{center}
\vspace{1cm}

\begin{abstract}
Characteristics of Abelian dyon - fermion bound system,
  parity - violating effects, a new series of energy spectra,
 effects related to the non - vanishing electric dipole moment,
 feature of  spin orientation  etc,
are analyzed and compared with hydrogen - like atom.
These analyses explore  possibility of a new approach of searching for
 dyons under bound condition.

\end{abstract}

\begin{flushleft}
${^\S}$ E-mail address:  jzzhang@physik.uni-kl.de  \\
\hspace{3.1cm} jzzhangw@online.sh.cn
\end{flushleft}
\clearpage
Since Dirac studied the problem of  quantum mechanics of a particle in
 presence of a magnetic monopole \cite{Dirac}, 
monopoles have been one of the interesting topics concerned in physics 
\cite{Schwinger}-\cite{HK}.
Their existence has been involved in  explanation of  phenomenon
 of electric charge quantization.
 Production of super-heavy monopoles or dyons (i.e. both electric 
and magnetic charged) in the early Universe 
is predicted in unified theories of strong and electroweak interactions 
\cite{Preskill} and 
its detection is one of  few experimental handles for these theories.

Because Montonen-Olive duality conjecture \cite{MO} which is 
manifestation of classical electromagnetic duality in some spontaneously
broken gauge theories, 
and its extension to an SL(2,{\bf \tt Z})
duality conjecture \cite{HT}
suggested by  Witten effect \cite{Witten79},
plays important roles in  recent developments of superstring theories
during the last few years, 
monopoles have been receiving renewed attention.

So far  search for free monopoles (or dyons) and for ones trapped
in bulk matter (meteorites, schist, ferromanganese nodules, iron ores
and others) has turned up negative.
A summary of experiments can be found in review papers
\cite{Giacomelli,Perl}.
In various experimental  schemes monopoles were assumed to have different
properties \cite{Giacomelli,Jeon,Perl}.
Some of  assumptions involved are
(i) electromagnetic induction, 
(ii) energy losses,
(iii) scintillation signature,
(iv) catalysis of proton decay, and
(v) trapping and extraction. Monopoles could be trapped in ferromagnetic
domains by an image force of  order 10 eV/{\AA}.
Trapped monopoles are supposed to be wrecked out of  material by large 
magnetic force.

There are several difficulties in searching for free monopoles.
First, we do not know how small the monopole flux is (according 
to the Parkker limit the up bound  is
$\phi < 3 \times 10^{-9} cm^{-2} y r^{-1}$ \cite{Parkker}),
so we do not know that in order to record a event how long we
have to wait.
Second,  estimation of monopole masses is model-dependent,
for example, masses of classical monopoles are about  order
$10-10^2$ GeV,
and masses of super-heavy monopoles in grand unified theories 
are about order $10^{16}$ GeV.
But we are ignorant of their definite values.
Specially we do not know whether masses of monopoles are within 
the energy region which 
can be reached by accelerator experiments in the near future.

In view of the fact whether monopoles (dyons) exist or not is 
important,
we may as well try to explore other means to find their existence.
If  monopoles ( dyons ) were produced in the early Universe, 
they would like to form bound states with charged fermions and 
remain in the present Universe.
In this letter we examine some detailed properties of dyon-fermion 
bound system, 
including their parity - violating effects, a new series of energy
spectra, trapping phenomenon by an inhomogeneous electric field
through a non - vanishing electric dipole moment, 
feature of the spin orientation, 
compare this system with hydrogen - like atom,
 and suggest a number of experiments to detect them. 

\vspace{0.5cm}
{\bf Charge of Dyon}. The Dirac quantization condition only
determines possible values $z$ of charge of fermion. 
Charge $z_d$ of dyon  is a free parameter which is not
determined by the Dirac quantization condition. 
In order to quantitatively analyze  dyon - fermion  bound system, 
values $z_d$ should be correctly determined. 
Consider two dyons with electric and magnetic charges, respectively, 
$(q = z_d e \hbar ,g)$ and 
$(q^{\prime} = z^{\prime}_d e \hbar , g^{\prime})$. 
The Zwanziger-Schwinger quantization condition \cite{Schwinger}
$qg^{\prime} - q^{\prime} g = 2\pi n, \, (n = 0, \pm 1, \pm 2,...)$
determines only difference between  electric charges of  dyons, 
$z_d-z^{\prime}_d = n$, 
but does not determine values of either $z_d$ or $z^{\prime}_d$. 
If CP is not violated,
under  CP transformation one determines that there are only two
mutually exclusive possibilities: 
either $z_d = n$ or $z_d = n + \frac12$ \cite{FO}.
 In  presence of a CP-violating term, electric charge of  
dyon explicitly depends on 
$\theta$ vacuum angle \cite{Witten79},
 $z_d = n + \frac{\theta}{2\pi}$.  Possibility $z_d = n + \frac12$ is
excluded, thus we have
\begin{equation}
z_d = n
\label{zd}
\end{equation}

In the following we review unusual properties of dyon - fermion
bound system in detail.
In order to provide experimental test all the  results 
are calculated according to new estimation of charge $z_d$ 
of dyon given by Eq.~(\ref{zd}).

\vspace{0.5cm}                                            
{\bf Parity violation.} In this system  spatial parity is
violated by  magnetic charge of
 dyon \cite{Goldhaber77,JZZ} because of  wrong transformation
property of magnetic
field $\vec{H}_D = g\vec{r}/r^3$ of dyon  under  space reflection
 $P$: $\vec{H}_D \to -\vec{H}_D$. 
Invariance of  Dirac equation in  external magnetic field under $P$
requires
that  vector potential $\vec{A}(\vec{x},t)$ transforms as
$P \vec{A}(\vec{x},t)P^{-1} = - \vec{A}(-\vec{x},t)$
which obviously contradicts  transformation of $\vec{H}_D$, unless one
artificially
changes  sign of magnetic charge $g$ under $P$. This parity 
violation leads to two effects:

(i) A modification of  selection rules of  electromagnetic
transition for
this system \cite{JZZ}. In  hydrogen - like atom, electric dipole
transitions are subject
to strict selection rules as regards  total angular momentum $j$ and
 parity
$P: |j^{\prime}-j| \le 1 \le j^{\prime}+j, \, P^{\prime}P = -1$, where
$j,P (j^{\prime}, P^{\prime})$ are total angular momentum and  
parity in  initial (final) state.
From $|j^{\prime}-j| \le 1$,  selection rules of total angular
momentum $j$ are
$\Delta j = 0, \pm 1$; but  parities of  initial and  final
states must be opposite,
thus  $\Delta j = 0$ transition is strictly forbidden. But for 
dyon - fermion  system,
parity is violated, thus $\Delta j = 0$ electric dipole transitions are
allowed.

 (ii) This system, different from  hydrogen - like atom, can possess a
non - vanishing electric dipole moment \cite{Goldhaber77}.

\vspace{0.4cm}
According to Ref.~\cite{KYG}  for a fixed $q$ there are three types 
of simultaneous eigensections of $\vec{J}^2, J_z$ and $H$ 
in dyon - fermion system,  types $\it A$ and $\it B$ 
 ($j\ge |q|+\frac12$), and type $\it C$ ($j=|q|-\frac12$).
Their eigensections are: 

\noindent for type $\it A$ ($j\ge |q|+\frac12$) 
\begin{displaymath}
\psi^{(1)}_{qnjm}=
\frac{1}{r}\left(\begin{array}{ccc}
h^{qnj}_1(r) \xi^{(1)}_{qjm}\\
-ih^{qnj}_2(r) \xi^{(2)}_{qjm}
\end{array}\right),
\end{displaymath}
for type $\it B$ ($j\ge |q|+\frac12$)
\begin{displaymath}
\psi^{(2)}_{qnjm}=
\frac{1}{r}\left(\begin{array}{ccc}
h^{qnj}_3(r) \xi^{(2)}_{qjm}\\
-ih^{qnj}_4(r) \xi^{(1)}_{qjm}
\end{array}\right),
\end{displaymath}
for type $\it C$ ($j=|q|-\frac12$) 
\begin{displaymath}
\psi^{(3)}_{qnjm}=
\frac{1}{r}\left(\begin{array}{ccc}
f^{qnj}(r) \xi^{(2)}_{qjm}\\
g^{qnj}(r) \xi^{(2)}_{qjm}
\end{array}\right),
\end{displaymath}
where
\begin{eqnarray}
&&\xi^{(1)}_{qjm}=c_{qj} \phi^{(1)}_{qjm}-s_{qj} \phi^{(2)}_{qjm}, \nonumber\\
&&\xi^{(2)}_{qjm}=s_{qj} \phi^{(1)}_{qjm}+c_{qj} \phi^{(2)}_{qjm}, \nonumber\\
&&c_{qj}=q[(2j+1+2q)^{\frac12}+(2j+1-2q)^{\frac12}]/[2|q|(2j+1)^{\frac12}], \nonumber\\
&&s_{qj}=q[(2j+1+2q)^{\frac12}-(2j+1-2q)^{\frac12}]/[2|q|(2j+1)^{\frac12}], 
\nonumber
\end{eqnarray}
\begin{displaymath}
\phi^{(1)}_{qjm}=
\left(\begin{array}{ccc}
(\frac{j+m}{2j})^{-\frac12} Y_{q,j-\frac12,m-\frac12}\\
(\frac{j-m}{2j})^{-\frac12} Y_{q,j-\frac12,m+\frac12}
\end{array}\right),
\end{displaymath}
\begin{displaymath}
\phi^{(2)}_{qjm}=
\left(\begin{array}{ccc}
-(\frac{j-m+1}{2j+2})^{-\frac12} Y_{q,j+\frac12,m-\frac12}\\
(\frac{j+m+1}{2j+2})^{-\frac12} Y_{q,j+\frac12,m+\frac12}
\end{array}\right).
\end{displaymath}
In the above $Y_{q,L,M}$ is monopole harmonic \cite{WY,Dray}.
Radial wave functions $R^{qnj}_i(\rho)=2ph^{qnj}_i(\rho)/\rho$ 
(i=1, 2, 3, 4) are \cite{ZQ}:

\noindent for type $\it A$
\begin{eqnarray}
R^{qnj}_{1,2}(\rho)=4p^2(M\pm E^D_{qnj})^{\frac12}A^{qnj}_1 e^{-\frac\rho2}
\rho^{\nu-1}\Bigl[F(-n, 2\nu+1,\rho) \nonumber\\
\mp \frac{n}{\mu+(\mu^2+n^2+2n\nu)^{\frac12}}F(-n+1, 2\nu+1,\rho)\Bigr]; 
\nonumber
\end{eqnarray}
for type $\it B$
\begin{eqnarray}
R^{qnj}_{3,4}(\rho)=4p^2(M\pm E^D_{qnj})^{\frac12}A^{qnj}_3 e^{-\frac\rho2}
\rho^{\nu-1}\Bigl[F(-n, 2\nu+1,\rho) \nonumber\\
\pm \frac{n}{\mu-(\mu^2+n^2+2n\nu)^{\frac12}}F(-n+1, 2\nu+1,\rho)\Bigr]. 
\nonumber
\end{eqnarray}
In the above $M$ is mass of fermion, 
$E^D_{qnj}$ is energy of dyon - fermion bound system 
 which is given by Eq.~(\ref{E-D}) below;
 $\rho=2pr,$ $p=[M^2-(E^D_{qnj})^2]^{\frac12};$ 
$n = 0,1,2,...$ is  radial quantum number;
$\nu =(\mu^2-\lambda^2)^{\frac12}> 0;$  
$\mu = [(j + \frac12)^2 - q^2]^{\frac12} > 0;$
$\lambda = zz_d e^2,$  
$z$ is electric charge of fermion, which is an integer; 
$j \ge |q| + \frac12;$
$q =zeg \not= 0$;  
Dirac quantization sets $eg = \frac{N}{2}, \; ( N =\pm 1, \pm 2, \pm 3,...)$
 (For the dyon case possibility $eg=0$ is excluded).
$F(a, b,\rho)$ is confluent hypergeometric function. 
$A^{qnj}_{1,3}$ are the normalization constants.
It is not necessary to show detailed structures of radial wave 
functions $f^{qnj}(\rho)$ and $g^{qnj}(\rho)$ of type $\it C$  for our purpose.

\vspace{0.5cm}
{\bf Energy Spectrum.}  Energy spectrum
of dyon - fermion bound system is \cite{LZ86} - \cite{JZZ}
\begin{equation}
\label{E-D}
E^D_{qnj} = M\left[ 1 + \frac{\lambda^2}{(n+\nu)^2} \right] ^{-\frac12}.
\end{equation}
 Spectrum (\ref{E-D}) is hydrogen - like, but there is delicate
 difference between spectra of a dyon - fermion bound system and spectrum 
of a ordinary hydrogen - like atom.
For hydrogen - like atom $j$ takes only half - integer. 
Total angular momentum of  dyon - fermion bound system includes a term 
$- q  \vec{r}/r$ contributed by monopole field, 
so $j$ takes half - integer as well as integer.

(i) When $q$ takes half - integer,
total angular momentum $j$ takes integer, 
leading to a new series of energy spectra that do not exist in 
ordinary hydrogen - like atom.

(ii) When $q$ takes integer, $j$ takes half - integer 
which is similar to the case of ordinary hydrogen - like atom. 
But compared with  energy level of the latter
\begin{equation}
E^H_{nj} = M \left[ 1 + \frac{(ze^2)^2}
{(n + [(j+\frac12)^2 -(ze^2)^2]^{\frac12})^2}
\right]^{-\frac12},   \nonumber
\end{equation}
 $E^D_{qnj}$ shifts down. Consider the case
$z = -1, \, z_d = 1, \, |q| = 1, \, j = \frac32,$ the shifted amount for 
$n=1$ energy level is
$(E^D-E^H)/M \sim 10^{-2}\alpha^2,$ 
where $\alpha$ is the fine-structure constant.
We also compare energy interval $\Delta E=E(n'=1)-E(n=0).$ 
The shifted  
$(\Delta E^D-\Delta E^H)/M $ is also at the order
$10^{-2}\alpha^2.$
Notice that these differences can be measured by present experiments.

For a Dirac monopole - fermion  bound system , there is 
LWP difficulty \cite{LWP} in  angular momentum state $j = |q| -
\frac12$. For a Daric dyon - fermion bound system    a new 
singularity occurs even in  angular momentum states
$j \ge |q| + \frac12$ when charge $z_d$ of dyon  exceeds a critical 
value $z^c_d$ \cite{LWZ}. 
In order to avoid this difficulty, one way is to introduce terms 
\cite{KYG,LWZ}
$-[\kappa ze/(2Mr^3)]\beta \vec{\Sigma}\cdot \vec{r}$
 and $-[\kappa\lambda/(2Mr^3)]\vec{\gamma} \cdot \vec{r}$.
Using the above wavefunctions $\psi^{(1,2)}_{qnjm}$ of  dyon - fermion 
bound system \cite{ZQ} we find that    
 energy shifts from term $\vec{\Sigma} \cdot \vec{r}$ vanishes
\begin{equation}
\label{dE1}
\Delta E^{(1)}_{qnj} = \langle -\frac{\kappa ze}{2Mr^3}\beta \vec{\Sigma}
\cdot \vec{r} \rangle_{qnjm} = 0. 
\end{equation}
This result is unlike to be accidental, behind it there should be a 
simple symmetry which needs to be explored. 
For  energy shift $\Delta E^{(2)}_{qnj}$ from term 
$\vec{\gamma} \cdot \vec{r}$
we consider $n = 0$ and the case of  dyon carrying one Dirac unit of 
pole strength, $|q| = \frac12$, thus $j = |q|+ \frac12 = 1;$ 
take $z = -1, \, z_d = 1,$ we obtain
\begin{eqnarray}
\label{dE2}
\Delta E^{(2)}_{\frac{1}{2}01} 
&=& \langle -\frac{\kappa z\lambda}{2Mr^3}
\vec{\gamma}\cdot \vec{r} \rangle_{\frac{1}{2}01m} \nonumber\\
&=&\frac{2\kappa\lambda^4M\Gamma(2\nu-1)}{\mu^3\Gamma(2\nu+1)}
=C_1 \kappa \alpha^4M.
\end{eqnarray}
where $C_1$ is a number of order 1. $\Delta E^{(2)}_{\frac{1}{2}01}$ 
can be neglected.
 The above results show that energy spectrum (\ref{E-D}) is quite
 accurate for a Daric dyon - fermion bound system with terms 
$\vec{\Sigma} \cdot \vec{r}$ and $\vec{\gamma} \cdot \vec{r}.$

For a Dirac dyon - fermion bound system coupled to general 
gravitational and electromagnetic fields their energy levels 
$\tilde{E}^D_{qnj}$ in the closed or open Robertson-Walker metric 
are \cite{LZ93}
\begin{equation}
\label{E-D1}
\tilde E^D_{qnj} = E^D_{qnj} \left[ 1 \pm 
\frac{\mu^2 \lambda^2(R_0/a_0)^2}
{6(\mu^2-\lambda^2)^{\frac12}[n+(\mu^2-\lambda^2)^{\frac12}] 
[n^2+\mu^2+2n(\mu^2-\lambda^2)^{\frac12}]}\right]
\end{equation}
where $R_0$ is the average radius of region of   dyon - fermion  system,
$a_0$ is the cosmological radius; 
the plus and  minus sign corresponds, respectively, to the closed and
 open space-time. 
After  epoch of recombination, the cosmological
radius $a_0(\tau)$ is about $10^{23}$ cm, 
$(R_0/a_0)^2 \sim 10^{-62}$ (if $R_0$ is about $10^{-8}$ cm). Thus
 correction of the curved space to  energy levels (\ref{E-D}) is
\begin{equation}
\label{dE-D1}
\frac{\tilde{E}^D_{qnj} - E^D_{qnj}}{ E^D_{qnj}} \sim
\frac{(R_0/a_0)^2}{(\mu^2-\lambda^2)^{\frac12}},
\end{equation}
which can be neglected. 
Only in the case of a large $z_d$ satisfied $\lambda \sim \mu$  
correction of the curved space  would become important.

\vspace{0.5cm}                                                       
{\bf Parity - violating Transition.}  matrix elements of the
$\Delta j = 0$ parity violation electric dipole transition  can be 
precisely estimated \cite{note1}. 
In  electric dipole approximation,
taking  transverse Coulomb gauge, Hamiltonian of this system is 
$H_I = - z e\vec{\alpha}\cdot \vec{\epsilon}A_0$,
where $\vec{\epsilon}$ and $A_0$ are, respectively,  polarization
vector and amplitude of  external electromagnetic field. 
For type $A$ we consider the case of
$q = \frac12, \, j = |q| + \frac12 = 1, \, n = 1, \, n^{\prime} = 0, 
z =-1, \, z_d = 1$.
Up to order $\alpha^2$, we have
\begin{equation}
\label{matrix1}
(H_I)^{(A)}_{qn^{\prime}jm,njm} = iC_2A_0 \epsilon_3m,
\end{equation}
\begin{equation}
\label{matrix2}
(H_I)^{(A)}_{qn^{\prime}j(m\pm 1);njm} =
iC_3A_0(\epsilon_1 \mp i \epsilon_2)(2 \pm m)^{\frac12}(1\mp
m)^{\frac12},
\end{equation}
where $C_2$ and $C_3$ are numbers of order $10^{-2}$.                   

For electric dipole transitions within type B states or between type 
$A$ and type $B$ states, results are similar to Eq.~(\ref{matrix1}) and 
Eq.~(\ref{matrix2}).

Transitions from type $A$ $(B)$ to type $C$ would presumably be crucial 
in identifying emissions from such a system.
 Electric dipole transition matrix elements from type $A$ to type $C$ 
are
\begin{eqnarray}
\label{matrix3}
(H_I)^{(A,C)}_{qn^{\prime}j^{\prime}m^{\prime};njm} &=&
iA_0 \delta_{j^{\prime},j-1}
\Bigl[ \delta_{m^{\prime},m-1}(\epsilon_1+i \epsilon_2)\frac{j+m}{2j}
\nonumber\\
&-&\delta_{m^{\prime},m+1}(\epsilon_1-i \epsilon_2)\frac{j-m}{2j} 
\nonumber\\
&+&\delta_{m^{\prime}m}\epsilon_3\frac{(j^2-m^2)^{\frac12}}{j} \Bigr]
\left(I^{(1)}_ {qn^{\prime}j^{\prime};nj}R_{qj}+
I^{(2)}_ {qn^{\prime}j^{\prime};nj}T_{qj}\right),    
\end{eqnarray}                                                   
where  $I^{(1)}_ {qn^{\prime}j^{\prime};nj}$ and 
$I^{(2)}_ {qn^{\prime}j^{\prime};nj}$ are radial integrals, 
$R_{qj}$ and $T_{qj}$  are numeral factors depending on $q$ and $j.$
 Eq.~(\ref{matrix3}) shows that  selection rule of  transition from 
type $A$ to type $C$ is $\Delta j = -1$; 
the $\Delta j = 0$ parity - violating transition is absent.
The result from type $(B)$ to type $C$ is similar to Eq.~(\ref{matrix3}).

\vspace{0.5cm}
{\bf Lyman Lines.}  Spectral series of transitions of this system
can be accurately estimated by Eq.~(\ref{E-D}). 
In the general case, $zz_de^2 \ll 1$. 
By Eq.~(\ref{E-D}), the photon wavelength 
$\lambda(q;n^{\prime}j^{\prime},nj)$
of  transition from  $(n^{\prime},j^{\prime})$ state to  $(n,j)$
state is
\begin{equation}
\label{lambda}
\lambda(q;n^{\prime}j^{\prime},nj) = \frac{4\pi}{M(zz_de^2)^2} 
\cdot \frac{(n^{\prime}+\mu^{\prime})^2 (n+\mu)^2}{(n^{\prime}
+\mu^{\prime})^2 - (n+\mu)^2}
\end{equation}
We consider the case $|q| = \frac12$. 
In this case $\mu =[j(j+1)]^{\frac12}, \, j = 1,2,3,...$.
For   dyon - electron system, $z = -1, \; z_d = 1$. 
We calculate the first Lyman line.
For transition  from 
$(n^{\prime} = 1, j^{\prime} = 1)$ state to  
$(n = 0, \, j = 1)$ state  is
\begin{equation}
\lambda_e(\frac12;11,01) = 2.8 \times 10^3\AA \,(\Delta j = 0 \, 
parity \, violation \, transition).
\nonumber
\end{equation}
 For transition  from 
$(n^{\prime} = 1, \, j^{\prime} = 2)$ state to 
$(n = 0, \,j = 1)$ state is
\begin{equation}
\lambda_e(\frac12;12,01) = 2.2 \times 10^3\AA \,(\Delta j = 1 \,parity 
\,conservation \,transition).
\nonumber
\end{equation}
For dyon - proton system, $z = 1, \, z_d = - 1$:
\begin{equation}
\lambda_p(\frac12;11,01) = 1.5 \AA \, (\Delta j = 0 \,parity \,
violation \, transition).
\nonumber
\end{equation}
\begin{equation}
\lambda_p(\frac12;12,01) = 1.3\AA \, (\Delta j = 1 \,parity \,
conservation \, transition).
\nonumber
\end{equation}
For   dyon - electron system the first Lyman lines are in the infrared
region.
For dyon - proton system the first Lyman lines are in the x-ray region.

\vspace{0.5cm}
{\bf Dipole Moment.}
Electric dipole moment $\vec{d}=e\vec{r}$ of this system can be 
represented by total angular moment 
$\vec{J}=\vec{r}\times(\vec{p}-ze\vec{A})+\frac12 \vec{\Sigma}-q\vec{r}/r$
as
$\vec{d}=e(-q\vec{r}+\frac12 \vec{\Sigma}\cdot\vec{r})\vec{J}/[j(j+1)].$
It is easy to show that  only its $z$ component  has
non - vanishing expectation value $\langle d_z \rangle_{qnjm}$
 in state $\psi^{(1,2)}_{qnjm}.$  
For the $n=0$ case, we have \cite{note2}
\begin{equation}
\label{moment}
\langle d_z \rangle^D_{qojm}=-\frac{eqm}{2j(j+1)}
\frac{\mu}{\lambda M\Gamma(2\nu+1)}
\end{equation}
Taking $q = \frac12, \, j = 1$, from Eq.~(\ref{moment}) it follows that 
\begin{equation}
\label{moment1}
\langle d_z \rangle^D_{\frac12 01m} \sim - C_4 (em/M),
\end{equation}
where $C_4$ is a number of order 10.

\vspace{0.5cm}
{\bf Spin Orientation.} For this system the expectation value 
$\vec{S}=\frac12 \vec{\Sigma}$ of spin of fermion is

\begin{equation}
\label{S}
\langle S_z \rangle^D_{qnjm} = \frac{m}{4j(j+1)}
\left( 1 + \frac{2\mu}{2j+1}\frac{E^D_{qnj}}{M}\right).
\end{equation}
Here $j$ and $n$ dependence in Eq.~(\ref{S}) is different from that in
hydrogen - like atom.
For hydrogen - like atom $ \langle S_z \rangle_{njlm}$ are
$\; \langle S_z \rangle_{nj(j+\frac12)m} = - m/\left[2(j+1)\right],$ \quad
 $\langle S_z \rangle_{nj(j-\frac12)m} = m/(2j)$.  In particular,
Eq.~(\ref{S}) depends on the radial quantum number $n$, but the latter
does not. The basic reason leading to the above difference is that in 
hydrogen - like atom spherical harmonic spinors $\Omega_{jlm}$ are 
eigenfunctions of $\vec{L}^2$, but in
dyon - fermion  bound system monopole spherical harmonic spinors
$\xi^{(1,2)}_{qjm}$ are not.

\vspace{0.5cm}
Based on the above examination of detailed properties of  Dirac 
dyon - fermion bound system, 
which are different from hydrogen - like atom, 
now we suggest the following experiments to detect them.

\vspace{0.5cm}
{\bf Analysis of Astronomical Spectrum.}  Approach of searching
for dyon - fermion bound systems,
compared with  approach of searching for free monopoles, shows advantage.
(i)  Superheavy dyon is treated as an external potential so that its
mass does not appear in  energy spectrum (\ref{E-D}).  
Spectrum (\ref{E-D}) is quite accurate,  corrections
from the term $\vec{\gamma} \cdot \vec{r}$ and  effect of  curved
space are completely negligible.
(ii) If dyons were  produced in plenty in the early Universe and formed
into bound states with charged fermions, 
radial electromagnetic spectra of these bound systems should 
 be recorded on astronomical observations during a long period.
There are some astronomical spectra recorded at the Kitt Peak 
Observatory 
which cannot be explained by atomic or molecular spectrum \cite{Fan}. 
We suggest to compare spectrum (\ref{E-D}) and  related Lyman lines 
with such unexplained astronomical spectrum.

\vspace{0.5cm}
{\bf Analysis of Trapped Dyon - Fermion Bound System.}
 Because  dyon - fermion
(electron, proton, etc.) bound system possesses a non - vanishing 
electric dipole moment $\langle d_z\rangle^D_{qnjm}$, 
it can be trapped by a well of  inhomogeneous electric field through
$- \langle \vec{d} \cdot \vec{E} \rangle^D_{qnjm}$.

(I) Residues in Ferromagnetic.
One way to obtain  electric dipole trap of  dusting material
of elementary ferromagnetic  in a trapping chamber is to
use a strongly focused laser beam with Gaussian intensity profile,
providing a field with an absolute maximum of laser intensity at 
the center of  focus. 
A laser trap relies on the force of an inhomogeneous electric field 
of a laser acting on the dipole moment of  dyon - fermion bound 
system \cite{note3}. 
In order to violate the condition of optical Earnshaw theorem one 
needs to properly switch the laser field on and off.
Because events are rare, we need high trapping density
\cite{note4}.

(II) Trapped Events from Cosmic Rays. In order to trap rare events 
from cosmic rays, 
we need to use  long duration static electric well with  inhomogeneous 
distribution.
In order to obtain a stable trap, it is necessary to violate the
 condition of static electric Earnshow theorem.

 In such trapping devices it is possible to check whether  trapped
objects are  dyon - fermion  bound system:

(i) One can observe their absorption spectrum and compare it with  
hydrogen - like atomic one according to Eq.~(\ref{E-D}).

(ii) Using  non - vanishing $\langle S_z \rangle^D_{qnjm}$, adding
 a strong magnetic field $\vec{\cal H}$ to orient the trapped 
system, one can examine  $n$ and $j$ dependence of
$\langle \vec{S} \cdot \vec{\cal H} \rangle_{qnjm}$
according to Eq.~(\ref{S}).

Investigation of  potential technical sensitivity of testing 
 dyon -fermion bound system in the above suggested experiments is out 
of this letter.

 Discovery of monopoles would have far-reaching consequences. 
Of course, any attempt to detect monopoles or dyons is a challenging 
enterprise. 
The reason is that if they remain in the present Universe they are 
surely rare.

\vspace{0.5cm}
This work has been supported by the Deutsche Forschungsgemeinschaft
(Germany). 
This work  has been 
also supported by the National Natural Science Foundation of China 
under Grant No.10074014 and by Shanghai Education Development 
Foundation.



\newpage
\vspace{0.5cm}
\begin{thebibliography}{99}
\bibitem{Dirac} P. A. M. Dirac, Proc. R. Soc. London, {\bf A133}, 60
(1931);
Phys. Rev. {\bf 74}, 817 (1948).
\bibitem{Schwinger} J. Schwinger, Phys. Rev.  {\bf 173}, 1536 (1968);
 Science,  {\bf 165}, 757 (1969);
D. Zwanziger, Phys. Rev.  {\bf 176}, 1489 (1968).
\bibitem{LWP} H.J. Lipkin, W. I. Weisberger and M. Peshkin, Ann. Phys.
(N.Y.)
  {\bf 53}, 203 (1969).
\bibitem{Parkker} E.N. Parkker, J. Astrophys.  {\bf 160}, 383 (1970);
B. C. Choudhary, hep-ex/9905023.
\bibitem{tHooft} G. 'tHooft, Nucl. Phys.  {\bf B79}, 276 (1974);
A.M. Polykov, Pis'ma ZETF,  {\bf 20}, 430 (1974).
[JETP Lett.  {\bf 20}, 174 (1974)].
\bibitem{WY} T.T. Wu and C.N. Yang, Nucl. Phys.  {\bf B107}, 365 (1976);
 Phys. REv.  {\bf D16}, 1018 (1977).
\bibitem{Jackiw} R. Jackiw and C. Rebbi, Phys. Rev. Lett.  {\bf 36}, 
1122 (1976).
\bibitem{KYG} Y. Kazama, C.N. Yang and A.S. Goldhaber, Phys. Rev.
   {\bf D15}, 2287 (1977).
\bibitem{Goldhaber77} A.S. Goldhaber, Phys. Rev.  {\bf D16}, 1815 (1977);
Y. Kazama, Phys. Rev.  {\bf D16}, 3078 (1977).
\bibitem{MO} N.S. Montonen and D. Olive, Phys. Lett. {\bf B72}, 117
(1977).
\bibitem{Preskill} J. P. Preskill, Phys. Rev. Lett.  {\bf 43}, 1365
(1979).
\bibitem{Witten79} E. Witten, Phys. Lett.  {\bf B86}, 283 (1979).
\bibitem{Giacomelli} G. Giacomelli, in Monopole '83, proceedings of the
 Ann Arbor meeting, Ann Arbor, Michigan, 1983, edited by J.L. Stone
(NATO Advanced Study Institutes, Series B: Physics Vol. III)
(Plenum, New York, 1984).
\bibitem{OW} P. Osland and T.T. Wu, Nucl. Phys.  {\bf B247}, 421 (1984);
  {\bf B247}, 450 (1984);  {\bf B256}, 13 (1985);  {\bf B256}, 32 (1985);
{\bf B256}, 4491 (1985);   {\bf B261}, 687 (1985).
 \bibitem{LWZ} Xin-zhou Lin, Ke-Lin Wang and Jian-zu Zhang, Phys. Lett.
  {\bf B148}, 89 (1984).
\bibitem{Dray} T. Dray, J. Math. Phys.  {\bf 26}, 1030 (1985);
 J. Math. Phys. {\bf 27}, 781 (1986).
\bibitem{LZ86} Xin-zhou Li and Jian-zu Zhang, Phys. Rev. {\bf D33}, 562
(1986).
\bibitem{FZQ} P.H. Frampton, Jian-zu Zhang and Yong-chang Qi, Phys. Rev.
  {\bf D40}, 3533 (1989).
\bibitem{ZQ} Jian-zu Zhang and Yong-chang Qi, J. Math. Phys.  {\bf 33},
1796 (1990).
\bibitem{JZZ} Jian-zu Zhang, Phys. Rev.  {\bf D41}, 1280 (1990).
\bibitem{LZ93} Xin-zhou Li and Jian-zu Zhang, J. Phys.  {\bf A26}, 4451
(1993).
\bibitem{HT} C. Hull and P. Townsend, Nucl. Phys.  {\bf B438}, 109 (1995);
E. Witten, Nucl. Phys.  {\bf B443}, 85 (1995).
\bibitem{FO} J. M. Figueroa-O'Farrill, Lectures on Duality, 1996
(unpublished).
\bibitem{Jeon} H. Jeon and M. J. Longo, Phys. Rev. Lett. {\bf 75}, 1443
 (1995); .
\bibitem{Lugo99} A. R. Lugo and F. A. Schaposnik, Phys. Lett. {\bf B467},
43
 (1999).
\bibitem{Tamaki} T. Tamaki, K. I. Maeda and T. Torii, Phys. Rev. {\bf
D60},
 104049 (1999).
\bibitem{Grandi} N. Grandi, R. L. Pakman, F. A. Schaposnik and G. Silva,
 Phys. Rev. {\bf D60}, 125002  (1999).
 (1999).
\bibitem{Goldhaber99} A. S. Goldhaber, Phys. Rep.  {\bf 315}, 83 (1999).
\bibitem{Kleihaus} B. Kleihaus and J. Kunz, Phys. Rev. Lett.  {\bf 85},
 2430 (2000).
\bibitem{Gamberg} L. Gamberg and K. A. Milton, Phys. Rev. {\bf D61},
075013 (2000).
\bibitem{Brihaye} Y. Brihaye, B. Hartmann and J. Kunz, Phys. Rev. {\bf
D62},
 044008 (2000).
\bibitem{Tripathy} P. K. Tripathy and F. A. Schaposnik, Phys. Lett.
 {\bf B472}, 89 (2000).
\bibitem{Lugo00} A. R. Lugo, E. F. Moreno and  F. A. Schaposnik, Phys.
Lett.
  {\bf B473}, 35 (2000).
\bibitem{Perl} M. L. Perl, hep-ex/0002001.
\bibitem{HK} B. Hartmann and B. Kleihaus and J. Kunz, Phys. Rev. Lett.
 {\bf 86}, 1422 (2001).
\bibitem{note1} In the context of monopole - fermion states for 
$j_{min} = 1q1-\frac12 > 0$, 
the $j_{min} \to j_{min}$ transitions have been also discussed by 
K. Olaussen, H.A. Olsen, P. Osland and I. Overbo, Nucl. Phys.  
{\bf B267}, 25 (1986).
 In the non-relativistic case  $\Delta j = 0$ transitions are discussed
 by E.A. Tolkachev, L.M. Tomil'chik, and Y.M. Shnir, J. Phys.
 {\bf G14}, 1 (1988).
\bibitem{note2} Kazama in Ref. \cite{Goldhaber77} estimated 
dominant term of $\vec{d}$ and $\vec{\mu}$ in a zero-energy bound 
state in the limit of very loosely bound approximation.
In our case besides a kinematic factor $m/[j(j+1)]$  
dynamical behaviors of
$\langle d_z \rangle_{qnjm}$ and $\langle \mu_z \rangle_{qnjm}$ 
are also obtained in detail.
\bibitem{Fan} Chang-Yun Fan, University of Arizona (private
communication).
\bibitem{note3} Because of  large mass and the perminent electric
dipole moment of  dyon - fermion bound system, 
 laser cooling and trapping of  events are easier than those 
of the neutral atom case. 
For this system maybe it is not so diffcult to break sub-Doppler 
cooling limit and subrecoil cooling limit to reach very low 
temperature.
 Recent report of the low temperature in three dimensional laser 
cooling beyond the single-photon recoil limit, see,  J. Lawall, 
S. Kulin, B. Sanbamea, N. Bigelow, M. Loduc
 and C. Cohen-Tannoudji, Phys. Rev. Lett.  {\bf 75}, 4194 (1995).
\bibitem{note4} In the neutral atom case, a far-off-resonance dipole 
force trap is used in  $R_b$ photoassociation experiments
to increase the trap density to $10^{12}$ cm$^{-3},$ see,
   J.R. Gardner, R.A. Cline, J.D. Miller, D.J. Heinzen,
H.M.J.M. Boesten and B.J. Verhaar, Phys. Rev. Lett.  {\bf 74}, 3764
(1995).

\end{thebibliography}
\end{document}